\documentclass[conference]{IEEEtran}

\usepackage{amsmath,amsfonts}
\usepackage{algorithmic}
\usepackage{graphicx}
\usepackage{textcomp}
\usepackage{xcolor}
\usepackage{tikz}
\usepackage{siunitx}
\usepackage{booktabs}

\begin{document}

\pdfpagewidth=8.5in
\pdfpageheight=11in

\newcommand{\iscasubmissionnumber}{215}
\pagenumbering{arabic}

\newcommand{\arch}[1]{Equilibria}
\newcommand{\meta}[1]{Meta}
\newcommand\kz[1]{{\color{blue}{{[KZ: #1]}}}}
\newcommand\abhishek[1]{{\color{orange}{{[Abhishek: #1]}}}}
\newcommand\neha[1]{{\color{red}{{[Neha: #1]}}}}
\newcommand\rev[1]{{\color{blue}{{#1}}}}
\newcommand\eliot[1]{{\color{cyan}{{[Eliot: #1]}}}}
\newcommand\dimitrios[1]{{\color{blue}{{Dimitrios:#1}}}}

\newcommand*\circled[1]{\tikz[baseline=(char.base)]{
            \node[shape=circle,draw,inner sep=0.4pt] (char) {#1};}}

\title{\arch{}: Fair Multi-Tenant CXL\\Memory Tiering At Scale}
\author{Kaiyang Zhao, Neha Gholkar$^1$, Hasan Maruf$^1$, Abhishek Dhanotia$^1$, Johannes Weiner$^1$, Gregory Price$^1$, Ning Sun$^1$, \\
Bhavya Dwivedi$^1$, Stuart Clark$^1$, Dimitrios Skarlatos \\ \\ Carnegie Mellon University, Meta$^1$}

\maketitle
\thispagestyle{plain}
\pagestyle{plain}

\begin{abstract}

Memory is becoming an ever-larger portion of system cost and power in datacenters. Memory expansion via Compute Express Link (CXL) provides an effective way to provide additional memory at lower cost and power.
However, making effective use of CXL memory requires software-level memory tiering support for most hyperscaler workloads. 

Many tiering solutions have been proposed. However, our observations from deploying CXL-enabled servers across a wide range of workloads at a hyperscaler reveal fundamental limitations in both prior approaches and the current Linux support. First, existing mechanisms lack multi-tenancy support, failing to robustly handle either stacked homogeneous or heterogeneous workloads. Second, their control planes offer limited flexibility in balancing memory across tiers, leading to fairness violations and uncontrolled performance variability. Finally, current systems provide insufficient observability, depriving operators of the visibility needed to diagnose performance pathologies at scale.

In this paper, we present \arch{}, the first operating system framework that directly tackles these challenges and enables fair, multi-tenant CXL memory tiering at datacenter scale. \arch{} provides per-container controls for memory fair-share allocation and fine-grained observability of tiered-memory usage and operations. It further enforces flexible, user-specified fairness policies through regulated promotion and demotion, and mitigates noisy-neighbor interference by suppressing thrashing.

We evaluate \arch{} in the fleet of a large hyperscaler using both production workloads and industry-standard benchmarks. Our results show that \arch{} consistently helps workloads meet their service level objectives (SLOs) while avoiding performance interference. Overall, \arch{} improves performance over the state-of-the-art Linux solution, TPP, by up to 52\% and 1.7$\times$ for production workloads and benchmarks, respectively. All \arch{} patches have been released to the Linux community.

\end{abstract}

\section{Introduction}

The rise in memory demands of datacenter applications has led to increased spending on memory in datacenters, including power consumption and capital expenditures~\cite{7838026,ramclouds}.
Memory can account for nearly a quarter of the rack power consumption and up to 44\% of the Total Cost of Ownership (TCO) of a typical compute server in \meta{}'s datacenters. 
Memory expansion based on CXL~\cite{cxl} has emerged as an effective way to mitigate the cost and power challenges. CXL is a cache-coherent interconnect that can attach memory backed by a multitude of technologies to the host~\cite{tpp,10.1145/3492321.3519556,10.5555/3386691.3386708,10.1145/3538643.3539745}. 
At \meta{}, recycled older-generation DRAM is connected to the host via CXL and presented as a NUMA node, providing 20--30\% of a server's total memory capacity, yielding significant cost savings.
We refer to CPU-attached DRAM as \textit{local memory} and CXL-attached DRAM as \textit{CXL memory}, and they are in two separate NUMA nodes.

Despite its potential benefits, CXL memory introduces higher access latency, making software support essential to prevent performance degradation. While applications can, in principle, explicitly manage allocations across NUMA nodes, our interactions with multiple product teams within and outside \meta{} reveal a strong preference for transparent memory-tiering mechanisms that abstract away this complexity from applications.
At the same time, datacenters are increasingly adopting multi-tenancy, where large servers are shared by multiple colocated workloads. 
Multi-tenancy is a highly effective strategy for maximizing the utilization of expensive hardware and improving performance per TCO. 
In practice, workloads often encounter bottlenecks in specific resources--such as CPU, network, or memory--leaving other resources underutilized. By colocating workloads with complementary resource demands, multi-tenancy enables more balanced and efficient resource utilization. 
This approach not only boosts the performance of individual compute nodes and reduces the number of servers required (thereby lowering power consumption), but also enhances overall fleet efficiency, ultimately improving cost-effectiveness and scalability across datacenters.

Prior work~\cite{tpp,pond,memtis,tmts,alto,nimble,nomad,flat2lm,colloid,10032695,CoAXIAL,StarNUMA,sun2023demystifying,mtm,arif2024application,vtmm,thermostat} has simplified the use of CXL memory through memory tiering, where hotter pages are transparently migrated to local memory and colder pages to CXL memory. 
However, multi-tenancy introduces new challenges for deploying CXL memory at scale that have not been identified or examined by prior work. In this paper, we show for the first time that these issues fundamentally limit the effectiveness and fairness of CXL memory tiering, and we systematically uncover and characterize these challenges using insights from real production deployments at \meta{}.

First, hyperscale services have diverse requirements for performance guarantees and service level objectives (SLOs), which complicates the decision to allow them to run on hosts with CXL memory due to performance uncertainties.
Existing memory tiering solutions lack multi-tenancy support and cannot guarantee fair memory sharing across tiers among colocated tenants. 
They primarily optimize for system-level metrics such as putting the system-wide hottest pages into local memory, ignoring fairness between colocated workloads. 
Even when they do consider multi-tenancy, they rely on coarse categorizations~\cite{tmts,pond} of workloads based on their importance such as \textit{critical/best-effort}. 
This ignores the problem that colocated "critical" workloads need to meet their own performance SLOs but still suffer from noisy neighbor effects. In fact, at \meta{}, the vast majority of multi-tenancy setups are multiple instances of the same workload with SLO requirements stacked together. 
This results in performance variability across containers and leads to severe SLO violations. 
For example, when two workloads with different memory access patterns are colocated, one may receive the majority of the local memory while the other receives very little under a conventional tiering scheme.
For another example, in a multi-tenant environment where tenants dynamically arrive and depart on hosts, workloads that launch later are disadvantaged, as they are forced to rely more heavily on CXL memory after local memory has been exhausted. 
As we will demonstrate in Section~\ref{sec:motivation} and \ref{sec:eval}, the lack of a fairness mechanism leads to performance losses of up to 65\% for \meta{}'s production workload \textit{Cache}, making CXL memory difficult to deploy for production workloads.

Second, operations in memory tiering such as demotion, promotion, setting up NUMA hint faults~\cite{hintfault,thermostat}, hint faults triggering, and others, are not virtualized across tenants (through containers or VMs). This makes it impossible to monitor the performance and tune the behavior of memory-tiering for colocated workloads, which is essential in today's datacenters that run first-party workloads.

Finally, most existing solutions are not thrashing-aware. We find that if a workload is thrashing between local and CXL memory, it can cause serious interference to colocated workloads and slow down the entire system. 
Therefore, detecting and mitigating thrashing is a prerequisite for a production-grade multi-tenant memory tiering solution.
All of these issues make CXL memory adoption in multi-tenant environments challenging, and new solutions are needed to pave the way for CXL memory in datacenters.

We build upon our learnings from deploying CXL memory in production and propose \arch{} for fair multi-tenant CXL memory tiering in datacenters. 
\arch{} is a solution in the OS that has three main components: 
(a) multi-tenancy-aware controls and observability for tiered memory -- interfaces that enable users to configure memory placement policies for distributing tiered memory among multiple tenants and monitoring tired memory utilization and operations per tenant;
(b) demotion and promotion regulations that apply the user-specified fair memory placement and meet the SLOs of colocated application; and 
(c) thrashing mitigation that reduces the adverse effects of a thrashing workload to the whole system.

We make the following contributions in this paper:
\begin{enumerate}
\item We share our learnings from deploying CXL memory in production in \meta{}'s datacenters.
They include the latest trends of memory hardware evolution, bandwidth, application adaptation,
as well as the acute fairness issues that are exposed when multi-tenancy is added to the equation.
To our best knowledge, we are the first to share observations from a real-world CXL memory deployment in datacenters with multi-tenancy enabled.

\item We design and implement \arch{} that provides per-container tiering observability, fair share controls and thrashing mitigation that help colocated workloads meet their performance SLOs.

\item We evaluate \arch{} with benchmarks and \meta{}'s production workloads. 
\arch{} meets the SLOs of colocated applications and avoids performance anomalies, providing up to 52\% and 1.7$\times$ performance improvements across production and benchmarks respectively over baseline Linux/TPP~\cite{tpp}. 

\item \arch{} is being deployed to the entire fleet of \meta{} and will unlock significant memory savings. We have shared \arch{} patches with the Linux community and a subset of patches have already been accepted upstream.
\end{enumerate}

\section{Background}

\subsection{Memory Overheads in Datacenters and CXL Memory}

\begin{figure}[t]
\centering
\includegraphics[width=0.9\columnwidth]{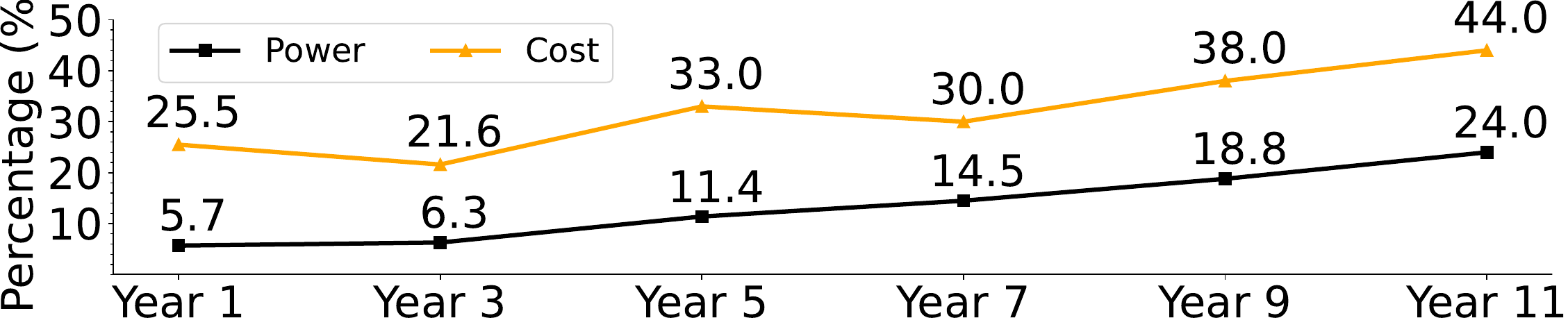}
\vspace{-2mm}
\caption{Memory as a \% of rack TCO and power at \meta{}'s datacenters.}
\vspace{-4mm}
\label{fig:memory-cost}
\end{figure}

Memory costs and power overheads have become a significant challenge in designing efficient systems for datacenters. Figure~\ref{fig:memory-cost} shows the contribution of memory towards the overall rack-level cost and power for one compute server types at \meta{}. Over the last decade, memory has grown to be one of the biggest contributors towards rack-level capital expenditures (CapEx) and operating expenses (OpEx)~\cite{pond,tpp}. 
The reasons for this trend are the following. First, workloads are dealing with more data, hence memory bandwidth and capacity requirements per unit of work have grown over time. Furthermore, memory technologies have not been scaling at the same rate as the other hardware components. Finally, some SoC architectures and packaging techniques (such as distributed last-level cache) lead to more cache misses and amplify memory traffic, which in turn requires provisioning more memory channels to meet bandwidth requirements and leads to higher memory power and cost.

Compute Express Link (CXL) presents a promising approach to addressing this challenge, offering greater flexibility in terms of capacity, cost, and performance characteristics. CXL is a cache-coherent interconnect that enables various memory technologies, including DDR, LPDDR, NAND, and persistent memory. CXL memory is attached to a host system, where it is presented as memory-only NUMA domains.
Notably, the DIMMs can be of different generations and potentially recycled from older servers, effectively reducing the carbon footprint of datacenters~\cite{greensku,10.1145/3604930.3605714}. Overall, CXL memory reduces overall system costs by supplying part of the host memory capacity at lower costs.

\subsection{Software Support for Memory Tiering}

CXL memory comes with limitations that must be addressed. 
CXL memory has increased access latency compared to CPU-attached DRAM, which can negatively impact application performance when significant portions of memory accesses are to CXL memory. 
This means that software support is required to effectively use CXL memory without compromising application performance. 
While applications could theoretically manage memory allocations across memory tiers explicitly, determining which data should reside in which memory tier creates substantial complexities that developers are not prepared to address in application code. 
Most tiered memory systems, therefore, rely on support in the OS that automatically migrates hot pages to the faster tier and colder pages to the slower tier.
Without such OS-level support, CXL memory would be substantially harder to deploy at hyperscale.

\subsection{Containers for Multi-Tenancy}

Datacenters have been moving towards larger servers in terms of cores and memory, which are multiplexed to support multiple tenants using containers or virtual machines (VMs).
In fact, containers have become the de facto standard for continuously developing, testing, distributing, and deploying applications at hyperscale.
Some operators choose to have an additional virtualization layer where a physical server is divided into VMs; nonetheless, containers still run inside those VMs.
Such containerization of application deployment has been shown to reduce the cost of running a modern datacenter for the following reasons. First, larger servers have lower TCO per unit of resources due to recent hardware trends~\cite{6290314}. Second, having one application use up the resources of an entire server is difficult, whereas containers allow multiple workloads to better fit the shape of the server, reducing stranded resources.
At \meta{}, multiple containers typically run side by side on the same physical server. 
Therefore, any CXL memory deployment must have full support for multi-tenant deployment with containers and provide predictable performance for each container.

Containers on Linux are built on top of 
control groups (cgroups) that enforce resource limitations, preventing individual containers from monopolizing the system. 
Linux tracks a container's memory usage and operations via cgroups~\cite{cgroups} and maintains metrics about its total memory usage and memory management operations applied to it, such as the number of page faults, reclamation, and compaction.
At \meta{}, host agents collect per-container metrics from cgroups to monitor workloads' performance, while a container orchestration tool sets the resource limits of each container using cgroups.
Therefore, \arch{} must provide container-level observability and configurability of memory tiering so that the existing container monitoring and orchestration stack continues to function.

Note that colocated VMs suffer from similar fairness issues as containers.
Since VMs launched on Linux can also be managed by cgroups, \arch{} similarly applies to VM deployments in datacenters.

\section{Learnings from Deploying CXL in Production}
\label{sec:motivation}

While CXL memory offers significant potential for reducing memory costs in datacenters, its practical deployment, especially in large-scale, multi-tenant environments, reveals several key challenges that must be addressed by the system software. 
In this section, we discuss the challenges we have seen from the initial stages of deploying CXL memory in \meta{}'s datacenters.

\subsection{Real vs. Emulated CXL Memory}

\begin{figure}[ht]
\centering
\includegraphics[width=0.95\columnwidth]{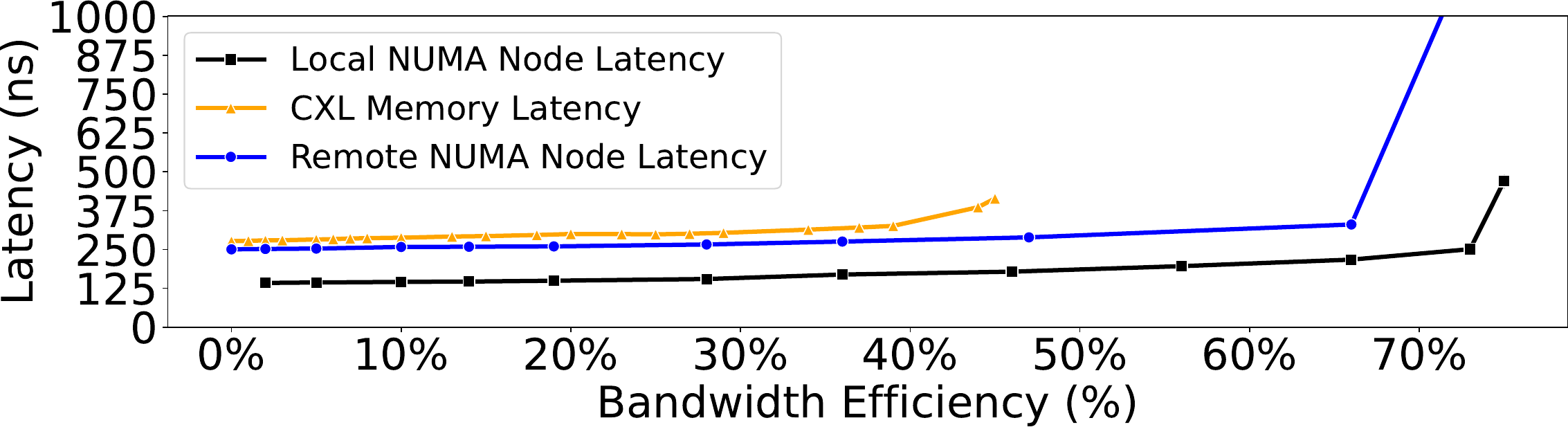}
\vspace{-3mm}
\caption{Latency versus bandwidth for local, remote, and CXL memory.}
\label{fig:cxl-latency-vs-bandwidth}
\vspace{-3mm}
\end{figure}

Prior work in CXL memory has primarily relied on emulating CXL memory by disabling CPU cores on multi-NUMA node machines.
However, we noticed that emulated CXL memory differs from real CXL memory in the following performance characteristics.
Figure~\ref{fig:cxl-latency-vs-bandwidth} depicts the latency versus bandwidth for
local, remote (across NUMA-node), and CXL-attached memory. Real CXL memory has higher idle latency (by 27ns) and higher loaded latency (by 37ns at 30\% load) compared to emulated CXL memory, 
while the achievable peak bandwidth efficiency is lower. 
This implies that workloads running on a real CXL system will generally be more sensitive to the efficacy of memory tiering than emulated systems might suggest, and highlights the importance of evaluating memory tiering solutions on real hardware.

\vspace{-2mm}

\subsection{CXL Bandwidth Use is Low for Capacity-Bound Workloads}
When running memory capacity-sensitive or capacity-bound workloads, we observe that most applications have a significant amount of cold memory that can be placed in CXL memory without compromising application performance. 
Table~\ref{tab:capacity-exp} shows data from three large representative production applications at \meta{} running on a system with 1TB of total memory (768GB local and 256GB over CXL). Each application uses a significant percentage of CXL memory capacity, but drives a very small amount of bandwidth on the CXL bus. 
This shows that we do not need to provision a large number of PCIe lanes to attach CXL memory devices to get the bulk of the benefits from CXL memory expansion. So, we continue to use a single x8 or x16 PCIe controller to connect CXL memory across several generations of servers.

\vspace{-2mm}
\begin{table}[ht]
  \centering
  \scriptsize
  \caption{Bandwidth usage of memory capacity-bound applications.}
  \begin{tabular}{|l|c|c|c|c|}
    \hline
    Application & Local Mem Used & CXL Mem Used & Local BW & CXL BW \\
    \hline
    App A & 718 GB & 101 GB & 157 GBps & 3 GBps \\
    \hline
    App B & 740 GB & 104 GB & 40 GBps & 1 GBps \\
    \hline
    App C & 508 GB & 151 GB & 41 GBps & 3 GBps \\
    \hline
  \end{tabular}
  \label{tab:capacity-exp}
  \vspace{-2mm}
\end{table}

\subsection{CXL Does Not Remove Memory Bandwidth Bottlenecks}

While CXL memory potentially offers both memory capacity and bandwidth benefits, an analysis of \meta{}'s production workloads and hardware indicates that additional memory bandwidth from CXL memory has limited practical use in current deployments. 
As shown in Table~\ref{tab:cxl-vs-ddr}, CXL memory provides only a small fraction of total system bandwidth, and the ratio of CXL to local memory bandwidth has not scaled over hardware generations. Furthermore, relying on DDR main memory bandwidth scaling is more power efficient than increasing the number of PCIe lanes to get additional memory bandwidth from CXL.

To illustrate this point further, Table~\ref{tab:bandwidth-exp} shows data from a memory bandwidth-bound production workload while leveraging different interleaving ratios to place a portion of application memory on CXL and utilize the additional bandwidth for performance improvements. Overall, we only see up to a 2.5\% performance upside on this application on a system configured to provide 10\% of its total memory bandwidth on CXL (the Gen1 system from Table~\ref{tab:cxl-vs-ddr}).
Therefore, we focus on the capacity expansion use case rather than bandwidth improvements, providing significant savings by extending memory capacity at low costs through the reuse of older DRAM.

\vspace{-2mm}
\begin{table}[ht]
    \centering
    \scriptsize
    \caption{Bandwidth and capacity scaling of CXL and DDR memory.}
    \begin{tabular}{|c|c|c|c|c|}
        \hline
        Hardware & CXL-Mem & Local-Mem & CXL-Mem: & CXL-Mem: \\
         Gen & BW & BW & DRAM BW. & DRAM Cap. \\
        \hline
        Gen1, Year 1 & ~30GB/s & ~300 GB/s & 0.1 & 0.25 \\
        \hline
        Gen2, Year 3& ~60GB/s & ~620 GB/s & 0.1 & 0.33 \\
        \hline
        Gen3, Year 3& ~60GB/s & ~620 GB/s & 0.1 & 0.375 \\
        \hline
    \end{tabular}
    \label{tab:cxl-vs-ddr}
    \vspace{-4mm}
\end{table}

\vspace{-2mm}
\begin{table}[ht]
  \centering
  \scriptsize
  \caption{Applications perf gains from more CXL memory bandwidth.}
  \begin{tabular}{|l|c|c|c|c|}
    \hline
    Interleaving & Local BW \% & CXL BW \% & Relative Perf \\
    \hline
    1:0 & 59\% & 0\% & 1.00 \\
    \hline
    2:1 & 31\% & 48\% & 0.97 \\
    \hline
    3:1 & 46\% & 25\% & 1.025 \\
    \hline
    4:1 & 48\% & 21\% & 1.021 \\
    \hline
  \end{tabular}
  \label{tab:bandwidth-exp}
  \vspace{-2mm}
\end{table}

\subsection{Most Applications Require Transparent Tiering}

Achieving broad adoption of CXL memory for capacity expansion hinges on minimizing the burden on application developers. 
In working with owners of major workloads at \meta{}, we observed a strong reluctance to implement custom memory management logic within their applications. 
Determining which data is hot or cold and managing its placement across memory tiers adds significant complexity, especially given the diverse server hardware generations and types on which these applications must run.
Consequently, we focus on transparent memory tiering mechanisms implemented at the OS level instead of pursuing per-application support. 

\subsection{Hardware Support for Tiering is Insufficient}

Ideally, hardware might assist with transparent tiering~\cite{10.1145/2749469.2750383,10.1145/3123939.3124555,8048924,10.1145/3368826.3377922,10.1145/3126908.3126923,ibs,pebs}, but current processor architectures and memory hardware have not evolved to adequately support memory tiering or address many of the existing limitations.
For instance, Intel's Two-Level Memory (2LM)~\cite{flat2lm,pmem2lm} provides hardware-managed tiering, but has crucial limitations. First, it requires a direct-mapped DRAM region as a cache for the CXL memory, necessitating a 1:1 hardware-managed local memory to CXL memory ratio. 
Secondly, it is not application-aware and cannot apply per-app policies.
Similarly, while CXL memory or DRAM could theoretically incorporate hotness monitoring units and autonomously perform memory tiering, such features are not widely implemented.

Given these hardware limitations, we build upon the software support in the OS to manage CXL tiered memory and ensure fairness under multi-tenancy.
In the future, we envision targeted hardware support for page migration and hotness tracking that works together with the OS to manage CXL memory with less overhead and more precision.

\subsection{Multi-Tenancy and Fairness}

\begin{figure}[t]
\centering
\includegraphics[width=0.9\columnwidth, trim = 30mm 22mm 22mm 22mm]{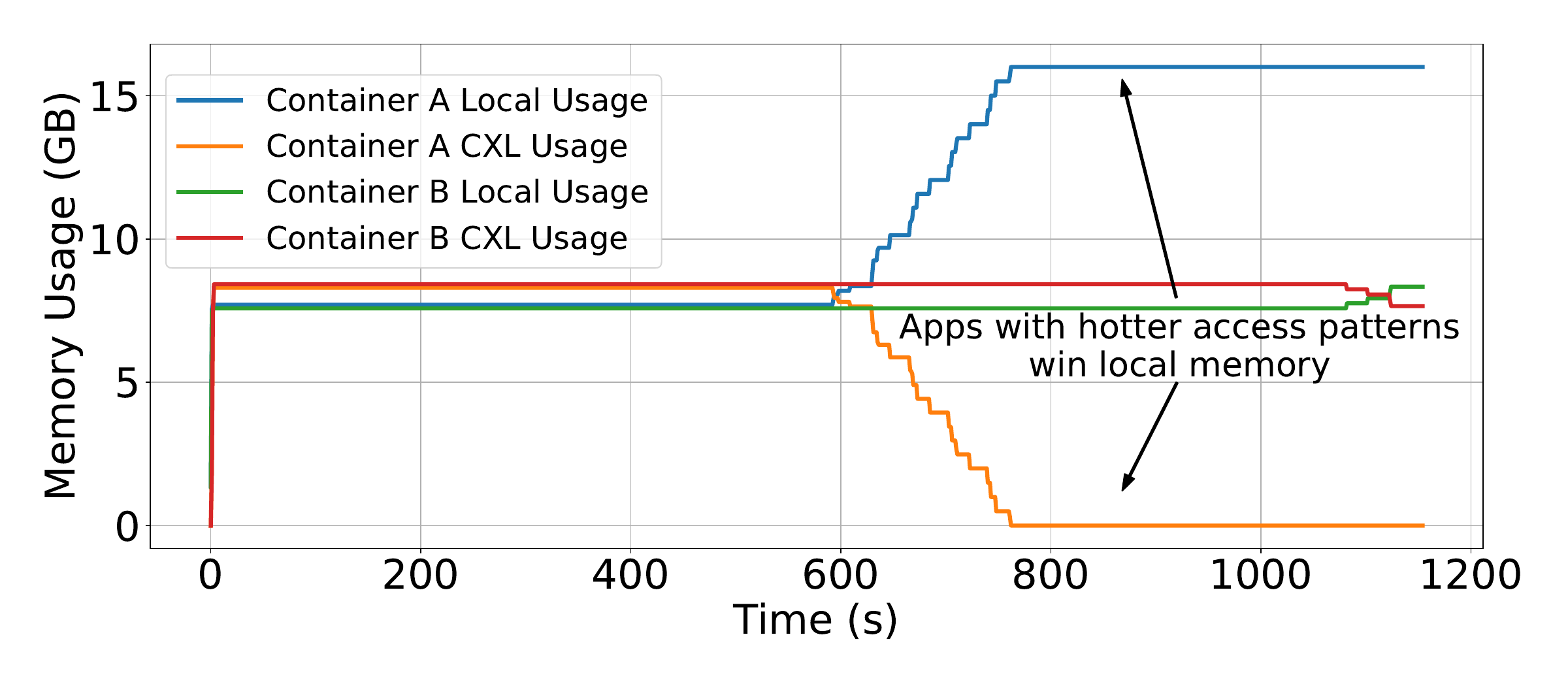}
\caption{Container A with hotter access patterns gets all of its footprint in local memory, whereas Container B gets only about half. }
\vspace{-4mm}
\label{fig:motivation-benchmark-1}
\end{figure}

Existing tired memory systems primarily make system-level decisions such as putting the system-wide hottest pages into local memory, ignoring fairness between colocated workloads. 
This can cause applications with colder access patterns to suffer.
To demonstrate this point, we used two instances (Container A, B) of a memory streaming microbenchmark that accesses memory at configurable hotness levels, where hotness is defined as the frequency of repeated memory access on a block of memory. 
As shown in Figure~\ref{fig:motivation-benchmark-1}, 
Container A, which has twice the hotness of Container B, is able to eventually have all its memory footprint in local memory when local memory is contended. On the other hand, Container B only gets about half of its memory footprint in local memory, leading to a lower access throughput compared to Container A.

Furthermore, there are performance variations between workloads that exhibit the same access hotness solely due to their launch order. We launch two instances of the microbenchmark with the same hotness, and Container B is launched 30 seconds later than Container A. 
Container A allocates memory first and is able to get all local memory. When Container B starts, local memory is exhausted, so it spills to CXL memory. Then, the memory distribution remains mostly static with limited demotion and promotion activities for both Container A and B, but the performance of Container B is permanently impaired with 28\% lower access throughput than Container A because of its high CXL memory usage.
Even though Linux attempts to keep a small amount (controlled by the \textit{low/min watermarks}) of local memory free so that new allocations (assumed to be hot) have a higher chance of landing on local memory~\cite{tpp} before falling back to CXL memory, kernel background/kswapd reclaim can only free up local memory when there are inactive pages, and often cannot keep up with the speed of new allocations so they fall back to CXL memory.

Finally, if a workload is thrashing between local memory and CXL memory, existing memory tiering solutions do not limit its negative performance impacts on other colocated workloads. We launch the microbenchmark with a working set significantly larger than the capacity of local memory, causing heavy thrashing. We also run two other instances of the benchmark with non-thrashing access patterns in CXL memory. We observe that under baseline Linux, there are constant and heavy promotion and demotion activities attributed to the thrashing workload at over 100K pages per second, while the other two workloads are deprived of memory tiering management attention with less than 1/10th of the page migration activities than when they run alone, so they are stuck using mostly CXL memory and have 7\% lower throughput than when they run alone.
This demonstrates that without proper mitigation, a thrashing workload has global negative effects on tiered memory systems.

In summary, the lack of fairness when colocated workloads use CXL memory leads to performance variability, 
and any tiering solution must resolve the fairness issue in multi-tenant environments to be deployed in production datacenters.

\section{\arch{} Design}
\label{sec:design}

We propose \arch{}, a solution in the OS to provide fairness in multi-tenant systems with CXL-based tiered memory.
The goal of \arch{} is to transparently managed tiered memory in multi-tenant datacenters while still meeting the performance service level objectives (SLO) of workloads.

\subsection{Defining Fairness}

We design \arch{} to achieve fairness in multi-tenant memory tiering systems.
Fairness is defined as each colocated workload being able to run at a performance level that meets its SLO, regardless of when it is launched and with whom it is colocated.
Without a concept of fairness, as shown in Section~\ref{sec:motivation}, workloads that launch later may experience lower performance due to the exhaustion of fast local memory, and workloads that are colocated with workloads exhibiting intensive memory access patterns may be forced to use too much CXL memory and miss their SLOs.
The goal of \arch{} is to help workloads always achieve their performance targets even with diverse colocated workloads and varying job schedules.

To achieve fairness, \arch{} focuses on memory placement of colocated workloads.
This is because, for the vast majority of production workloads at \meta{}, ensuring that a certain ratio of the hottest pages are in fast local DRAM is sufficient to guarantee that the performance of a workload using tiered memory is at a target level.
We call it the fair share of local memory.
There are two benefits of the approach. First, determining the necessary ratio of local DRAM for each workload can be achieved through a few profiling runs of a single workload offline, without enumerating all possible combinations of colocated workloads.
Second, the other potentially contended resource---CXL memory bandwidth---is shown in Section~\ref{sec:motivation} to be not under contention for memory-capacity-bound workloads, and not having to specify more parameters for tiering simplifies job configurations.

The fair share of memory is subjective and depends on workload characteristics and SLOs. In some use cases, we may prefer to distribute local memory evenly, while in others we may prioritize applications that are sensitive to memory access latency. We designed \arch{} to be highly configurable, enabling service owners to implement various fairness policies tailored to their specific performance objectives.

\begin{figure}[t]
\centering
\includegraphics[width=0.9\columnwidth]{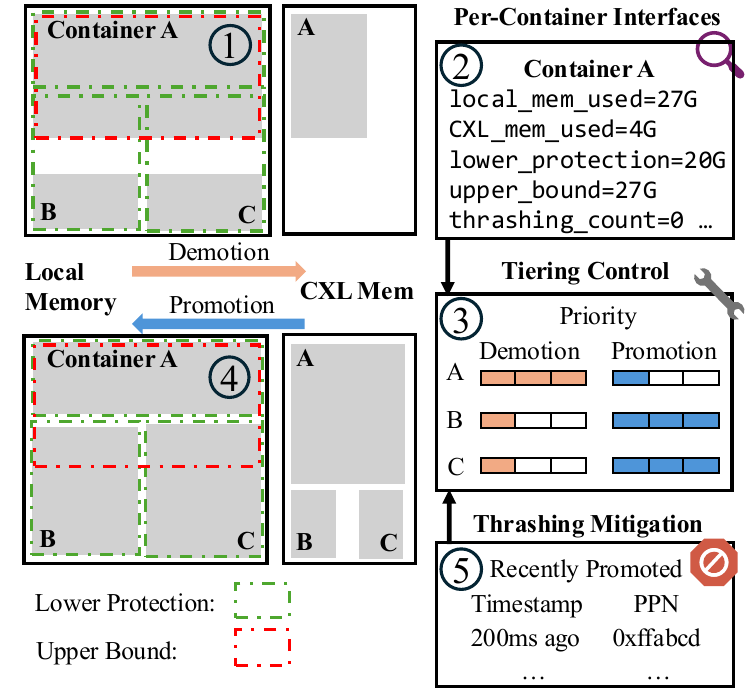}
\vspace{-2mm}
\caption{\arch{} Overview.}
\label{fig:design-overview}
\vspace{-5mm}
\end{figure}

\subsection{Overview}

\arch{} enables the fair sharing of memory through two parameters per container. The first parameter, \textit{lower protection}, is the amount of local memory protected from demotion or reclaim due to competing memory usage by other colocated containers when local memory is fully utilized.
The second, \textit{upper bound}, is the maximum amount of local memory a container can use before demotion reduces its local memory usage. 
They serve different purposes. The lower protection guarantees that a workload will get its fair share of memory regardless of the behavior of its neighbors, and avoids SLO violations. It still allows a container to opportunistically use available local memory exceeding its fair share, making local memory allocation work-conserving and enabling burstable performance. The upper bound, on the other hand, is a hard stop of a container's local memory usage, useful when such higher than usual performance is undesirable for capacity planning, or when free local memory should be reserved for new workloads to quickly ramp up.
By setting these parameters, \arch{} allows administrators to balance fairness, performance, and predictability in tiered memory systems.

Figure~\ref{fig:design-overview} gives a high-level overview of \arch{} and how it ensures memory placement fairness under multitenancy. 
In the beginning, Container B and C are using less than their local memory lower protection (green dotted boxes), allowing Container A to opportunistically use take up the unused local memory. However, A is stopped from using all free local memory by its upper bound (red dotted box) and has to spill 4GB to CXL memory (Step \circled{1}).
\arch{} monitors for each container the local and CXL memory usage and tiering activities, such as demotions, promotions, and reclamations. The information is exported to the userspace via cgroup interfaces for fleet performance monitoring. The local memory lower protection and upper bound are also set via the interfaces (Step \circled{2}).
After a while, the memory demands of Container B and C increase, and system local memory capacity is now contended. \arch{} starts background demotion to free up local memory for hot new allocations. In the process, it prioritizes demotion and deprioritizes promotion for Container A since it is exceeding its local memory lower protection; conversely, \arch{} deprioritizes demotion for Container B and C so that they are guaranteed to get their local memory lower protections (Step \circled{3}).
Soon, the local memory usage of containers converge to their designated lower protections under contention (Step \circled{4}).
Finally, \arch{} is always detecting when pages are frequently migrated between local and CXL memory, a sign that they are possibly thrashing due to unfavorable access patterns and excessively large working sets, and slowing down promotions for thrashing containers to reduce noisy neighbor effects (Step \circled{5}).

\subsection{Container-level Observability in Memory Tiering}

\arch{} extends the tracking of per-container memory usage to a new dimension: memory tier. To support the separate monitoring of local and CXL memory usage for each container, \arch{} adds new memory usage counters within the cgroup's data structure for each tier. \arch{} also adds new tiering activity counters for each container, covering demotions, promotions, attempted promotions, and other relevant events. When a page is allocated or released, the memory tier of the physical page is identified, and the relevant usage counters of the owning container are updated. Similarly, when a page is migrated, the relevant promotion or demotion counters of the container are updated. In this way, \arch{} ensures that crucial information about tiered memory usage and operations is maintained for each container to support memory placement fairness, and makes this data available to other external resource monitoring tools.

\subsection{Modulating Demotions}

\arch{} manages the rate of demotion for each container to realize the fair sharing of local memory.
\arch{} differentiates tiers of memory and accepts different memory lower protection and upper bounds for each tier in the hierarchy, whereas Linux does not consider memory from different tiers as needing separate resource control due to their distinct access performance.
When there is free local memory, containers are allowed to use more local memory than their lower protection. This approach improves the utilization of local memory, which offers lower latency and higher bandwidth compared to CXL memory, thereby optimizing the server's total throughput.

When local memory is fully utilized, \arch{} begins to assign priorities to containers for demotion.
Containers with local memory usage lower than their lower protection are exempted from demotions. Those exceeding their lower protection are targeted for demotions with intensity proportional to the overage. More specifically, if $n_{cgroup}$, $n_{protection}$, $n_{lru}$ are the container's local memory usage, lower protection and the size of a local memory LRU list of the container, respectively, the number of pages from the LRU list to be considered for demotion $d_{scan}$ is calculated as:
\begin{equation}
d_{scan} = \frac{n_{lru}(n_{cgroup}-n_{protection})}{n_{cgroup}}
\label{equation:dscan}
\end{equation}
After $d_{scan}$ is calculated for a container, it is passed to the reclaim algorithm in Linux. Finally, a set of least recently used pages owned by the container will be selected and demoted. The process is repeated for every container in the system to free up local memory.
Compared to Linux, \arch{} differentiates memory reclaim due to insufficient memory in the system from demotion due to limited local memory capacity, and applies different priorities based on the type of memory shortage.
Over the long term, the prioritization of demotion by \arch{} applies higher demotion pressure to containers that use more local memory than their fair share given their neighbor containers, pushing the local memory usage of each container in the system to a stable state as prescribed by the system administrator.

\arch{} enforces local memory upper bound by monitoring a container's memory usage whenever it allocates memory. Unlike the lower protection, a container cannot use more local memory than the amount specified in upper bound. As the usage approaches the upper bound, \arch{} initiates a kernel background thread to demote pages and decrease the container's local memory footprint, gently enforcing the limit with minimal disruption to the application's execution. If the container's local memory usage reaches or surpasses the limit during allocation, \arch{} requires the allocating thread to synchronously perform demotion before resuming userspace execution. This approach slows down future allocation requests for applications that try to rapidly allocate local memory without resorting to out-of-memory kills.

\subsection{Regulating Promotion}

To achieve a fair share of the local memory, promotion must work in synergy with demotion to transition the memory placement of containers to the desired state. \arch{} de-prioritizes promotion of containers that use more than their fair share of local memory, in concert with the prioritization of demotion for such containers.

\arch{} categorizes a container as ``promotion throttled'' in either of the two cases: if its local memory usage exceeds its lower protection and the local memory is fully utilized, or if its local memory usage is approaching or exceeds its local memory upper bound. \arch{} sets a flag for the container if either of these conditions is true. During the promotion candidate page scanning for a container marked as promotion throttled, \arch{} reduces the number of pages to scan for promotion candidates via the following calculation, where $p_{scan}$ is the size of the container's CXL memory footprint to consider for promotion, and $p_{base}$ is the unthrottled promotion scanning rate:
\begin{equation}
p_{scan} = p_{base} * \min(\frac{n_{protection}^4}{n_{cgroup}^4}, \frac{1}{16})
\end{equation}
In other words, promotion scanning is slowed down in proportion to the fourth power of the overage amount over a container's guaranteed lower share of local memory.
We choose the power function so that the throttling starts minor but intensifies ever faster with more overage, i.e., promotion scan is at 96\% of the normal scan size when the overage is 1\%, but at 10\% overage the promotion scanning is only 68\% of the base rate. 
The reason behind this is that a small overage is likely due to a burst of allocations that cause a temporary mismatch between the speed of demotion and allocation, which will subside on its own; on the other hand, a larger overage signals the inability of the demotion to keep up, necessitating a more drastic slowdown of promotion to more effectively pare back the container's local memory usage.
Notice that \arch{} still allows promotions to proceed when the local memory usage of a container exceeds its fair share because it may have very hot pages in CXL memory; otherwise, those pages may be stranded in CXL memory and excessively slow down its execution. By allowing a minimum of one-sixteenth of the base promotion scanning rate, \arch{} trades memory placement convergence speed for application performance.
In summary, the regulating of promotion speed occurs in conjunction with prioritized demotion for such containers to reduce their local memory footprint and ensure that demotion and promotion are not fighting each other. 

\subsection{Mitigating Thrashing}

\textit{thrashing} is when pages are repeatedly and rapidly promoted and demoted between local and CXL memory. It can occur when a workload's hot working set surpasses the local memory capacity, leading to pages being frequently promoted to local memory only to be demoted shortly thereafter to accommodate other hot pages, creating a cycle. 

Thrashing between memory tiers is detrimental to performance, diverting compute cycles and memory bandwidth from application execution. In a CXL tiered memory system, it is particularly undesirable, as direct access to CXL (albeit at higher latency) is more efficient than migrating pages to local memory only to demote them again before they are subsequently accessed. Furthermore, thrashing increases both CXL and local bandwidth loads, thereby increasing access latency for direct CXL accesses as well. For this reason, in multi-tenant environments a thrashing container can create noisy neighbor problems, increasing the access latency for neighbors and impacting their performance. Therefore, mitigating thrashing is an essential component of \arch{}.

\arch{} identifies thrashing containers and limits their promotion rates. 
\arch{} maintains lightweight tracking of recently promoted pages by sampling them in a fixed-size hash table that records the physical page number and timestamp right after a promotion. The sampling helps to reduce the memory consumption of the tracking. During demotion, if the page is found in the hash table and the elapsed time between promotion and demotion is lower than a threshold $t_{resident}$, the event is considered a thrashing occurrence. The container that owns the page will have its thrashing counter incremented.
\arch{} includes a background thread that wakes up every 5 seconds and performs the following task: first, it clears the hash table that tracks recently promoted pages, moving the thrashing detection to the next period. Second, it calculates the rate of change in the thrashing counter of each container over the last period. If the rate exceeds a threshold $r_{thrashing}$, the promotion rate for pages owned by the container will be halved for the next period. And if a previously thrashing container falls below $r_{thrashing}$, its promotion rate will be doubled for the next period until the normal promotion rate is restored. 

However, all thrashing containers are not the same. \arch{} monitors the memory statistics of containers to differentiate between those experiencing working set transitions (not yet reached steady state), and those that are thrashing in steady state. This distinction is crucial because we want to avoid limiting promotion rates for workloads undergoing working set transitions, as it could prolong their convergence and result in performance degradation.
To define the memory steady state for \meta{}'s services, we uniform-randomly sample 5000 servers from the pool of servers running five major services, and study the distribution of memory metrics in correlation to the services' health status.
We identified the following two metrics to be indicative of whether a container has reached a steady state: the rates of change of the number of active pages and the rate of page freeing -- when a container's both metric values are below thresholds, the container is considered in steady state.
\arch{} applies thrashing mitigation only to containers in steady state.

\subsection{Implementation}

We implemented \arch{} on top of the Linux kernel 6.11, using approximately 1,000 lines of code. 
\arch{} integrates with the cgroup memory controller to track and export local memory usage and tiering operation counts for containers, and provide knobs for setting the local memory lower protection and upper bound. This means that our solution works uniformly across containerized and virtualized environments. We have shared the patches with the Linux community and a subset of patches have been accepted upstream.

\arch{} is implemented in a modular way with upstreaming to Linux in mind, and the changes are contained in the cgroups, demotion, and promotion code of Linux, making it easier to test and upstream.
By adhering to the existing cgroup-based memory resource control and pursuing an upstream-first approach, we avoid ongoing code maintenance costs associated with designs that heavily deviate from how Linux currently manages memory.
Being built into the operating system also allows \arch{} to benefit the broadest set of applications in a transparent manner, while also avoiding the overhead of performing user-space-to-kernel communication. 

\section{Evaluation}
\label{sec:eval}

\subsection{Methodology}
\label{sec:methodology}

We performed experiments on two types of \meta{} servers equipped with CXL memory, denoted as small and large servers. 
Both are single-socket servers with CPUs of the AMD Turin generation. 
The large server features a local to CXL memory ratio of 3:1, including 768GB of local DDR5 memory and 256GB of DDR4 DRAM as CXL memory. It has an unloaded CXL memory access latency of 252ns. 
The small server has 256GB of local DDR5 memory and 64GB of DDR4 DRAM as CXL memory. 
The CXL memory expanders used in the evaluation are CXL 2.0 Type 3 devices, using recycled previously-decommissioned DDR4 memory.

We compare \arch{} against the upstreamed memory tiering support in Linux kernel, which is derived from TPP~\cite{tpp}. 
The servers have sufficient combined local and CXL memory capacity for colocated workloads, and swapping is disabled.
Although we do not provide a quantitative evaluation of the container-level observability of memory tiering provided by \arch{}, most of the evaluation and motivation studies in this paper are only possible because of it.

We evaluate three of \meta{}'s production workloads: Cache, Web, and Continuous Integration (CI), as well as two benchmarks from DCPerf~\cite{dcperf}: SparkBench and TaoBench.
\textit{Cache} is a distributed object cache that provides low latency access, defined consistency models, and common query semantics.
It is the foundation of many internal and user-facing workloads at \meta{}.
Cache receives replayed requests from production traces.
\textit{Web} is an application server that hosts many of \meta{}'s user-facing services. 
\textit{CI} automates the process of building and testing code changes before integrating them into a shared repository. 
Both Web and CI serve live requests in production.
\textit{DCPerf} is a suite of datacenter benchmarks that closely approximates major services in datacenters.

\subsection{Validating \arch{} }
\label{sec:eval-validation}

\begin{figure}[t]
\centering
\includegraphics[width=0.9\columnwidth, trim = 30mm 22mm 22mm 0mm]{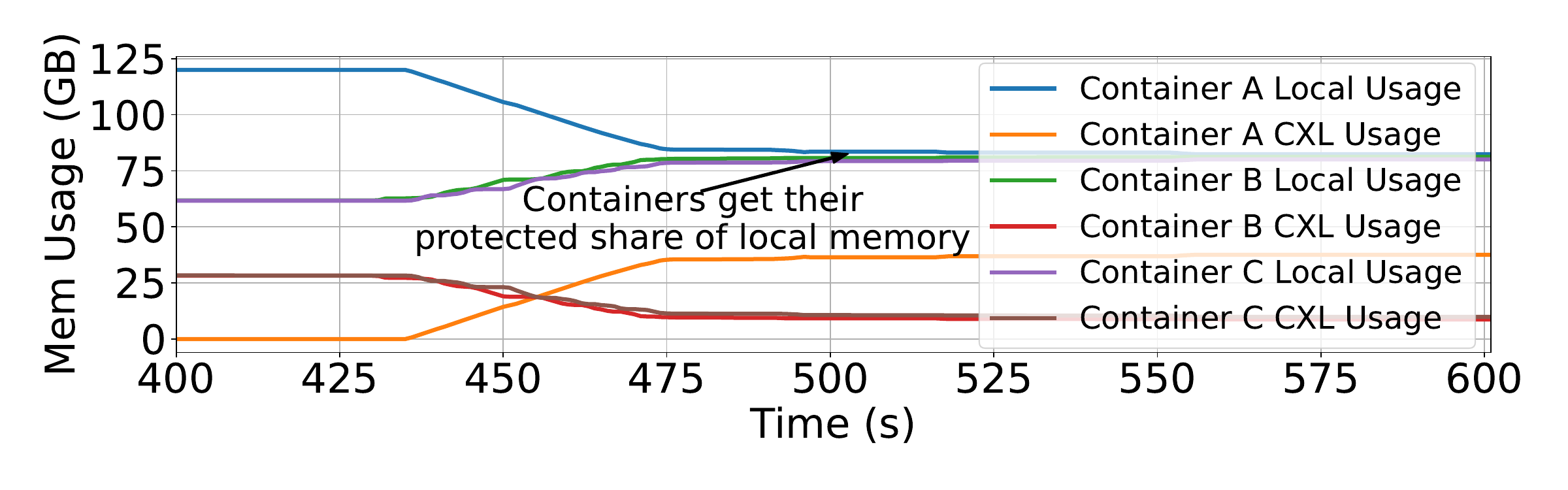}
\caption{Memory usage when containers exceed local memory lower protection.}
\label{fig:benchmark-2-memory}
\vspace{-4mm}
\end{figure}

In this section, we validate the functionality of \arch{}'s components. 
These experiments were conducted on a small server. 
We designed a microbenchmark that allocate a predetermined amount of memory and access it through sequential passes. To simulate a multi-tenant environment, we launched three instances of this workload concurrently, each running within its own container. 
We used microbenchmarks due to their deterministic behavior and ease of analysis, allowing for clear reasoning about the system's behavior.
We consider the following four conditions: whether there is local memory pressure, whether there is unused local memory protection that can be donated, whether the upper bound is set, and whether there is thrashing. 
In the following experiments, the local memory lower protection is set to 80GB. 
No limits are imposed on CXL memory usage for any containers.

\subsubsection{Local memory is preferred}

When local memory is not contended, each container is allowed to reside entirely in local memory, even if exceeding its lower protection value (80GB). To validate this behavior, we launched three microbenchmarks with respective footprints of 120GB, 40GB, and 40GB. Given that their combined footprint (200GB) is less than the available local memory capacity, no contention occurs. We observe that \arch{} allows all three containers to be fully resident in the local memory.
This behavior of \arch{} increases the system-wide utilization of local memory to get potentially higher total throughput out of a server.

\begin{figure}[t]
\centering
\includegraphics[width=0.9\columnwidth, trim = 30mm 22mm 22mm 0mm]{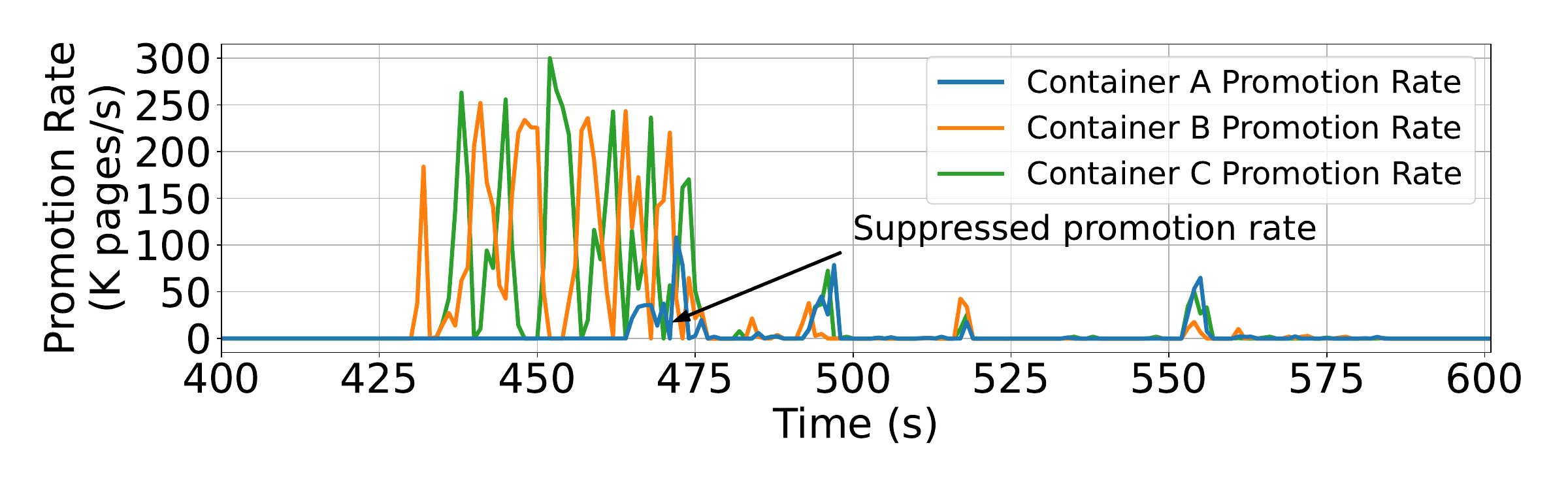}
\caption{Promotion activities when containers exceed local memory lower protection. Container A's promotion rate (blue line) is throttled.}
\label{fig:benchmark-2-promotion}
\vspace{-4mm}
\end{figure}

\subsubsection{Lower protection is enforced}

In this scenario, we launch three instances of the microbenchmark having memory footprints of 120GB, 90GB, and 90GB, respectively, each of which exceeds the per-container local memory lower protection. When the local memory is fully utilized and contended, each container's local memory footprint eventually converges to its lower protection value of 80GB. The remaining memory footprint for each container is then spilled over into CXL memory, with Container A having 40GB and Container B and C having 10GB each, as illustrated in Figure~\ref{fig:benchmark-2-memory}.

We also observed that Container A is demoted more aggressively compared to the other two containers, with a rate of up to 250K pages per second. This is due to Container A exceeding its local memory lower protection by 50\%, whereas Containers B and C only exceed it by 13\%. As a result, Container A is demoted more heavily to bring its memory usage down to its fair share.
Additionally, we note that the promotion rate of Container A is suppressed from time 425s to 475s, as shown in Figure~\ref{fig:benchmark-2-promotion}. Despite using more CXL memory and having more hot candidate pages in CXL for promotion, Container A's promotion rate is intentionally limited by \arch{}. This ensures that promotions do not interfere with converging to a steady state.

\subsubsection{Unused lower protection is donated}

We launch three microbenchmark instances with memory footprints of 120GB, 70GB, and 70GB. Container A is launched first and utilizes all available local memory before the other containers start. The memory usage of three containers converges over time.
Container A opportunistically uses 20GB more local memory than its lower protection, even when local memory is contended. 

This is because Container B and C donate their unused local memory share to Container A. In this case, where donation happens, \arch{} still ensures that B and C have all their protected local memory share and fully reside in local memory.
This experiment demonstrates the work-conserving nature of \arch{}'s lower protection mechanism. In production, a similar scenario can help workloads achieve higher performance when the system has free local memory.
Note that Containers B and C are completely protected from demotions because their memory usage falls under their local memory lower protection, and we find that only Container A experiences demotion activities.
This means that collocated workloads do not need to worry about degraded performance when their unused share of local memory is donated.

\subsubsection{Upper bound is enforced}

In this experiment, three microbenchmark instances with memory footprints of 120GB, 40GB, and 40GB are used, similar to the ``no local memory pressure'' case. However, Container A's local memory upper bound is set to 80GB. As a result, despite ample system-wide local memory availability, Container A is restricted from using more than 80GB of local memory, forcing it to spill 40GB into CXL memory.
This scenario represents scenarios where minimizing performance variability is paramount, and local memory should be reserved for new neighbors rather than being opportunistically utilized by existing workloads.

\subsubsection{Thrashing is mitigated}

In this experiment, we demonstrate the effectiveness of \arch{} in detecting and mitigating thrashing workloads, resulting in improved system-wide performance.
We tune the aforementioned microbenchmark to create a thrashing workload as follows: it allocates 100GB of memory and accesses it in blocks such that the memory is hot enough to trigger page promotions from CXL memory but not enough to re-access the promoted pages before they are demoted. The local memory upper bound for the container is set to 10GB.
We run two other microbenchmark instances with 50GB of memory footprint each, representing normal workloads that do not thrash. These workloads fit within their local memory lower protection.
We compare two scenarios: one in which thrashing mitigation is disabled and another in which it is enabled.
Without thrashing mitigation, the thrashing workload experiences aggressive page migrations (over 100K pages per second), causing interference with the other two normal workloads, which access memory at a rate of 290K pages per second.
However, when thrashing mitigation is enabled, \arch{} detects the thrashing behavior and reduces page migrations to thousands of pages per second, significantly decreasing system-wide memory management overhead. As a result, the colocated well-behaving programs can sustain access to memory at a rate of 310K pages per second, achieving a 7\% performance improvement.

\subsection{Evaluating \arch{} with Benchmarks}

\begin{figure}[t]
\centering
\includegraphics[width=0.9\columnwidth, trim = 30mm 22mm 22mm 0mm]{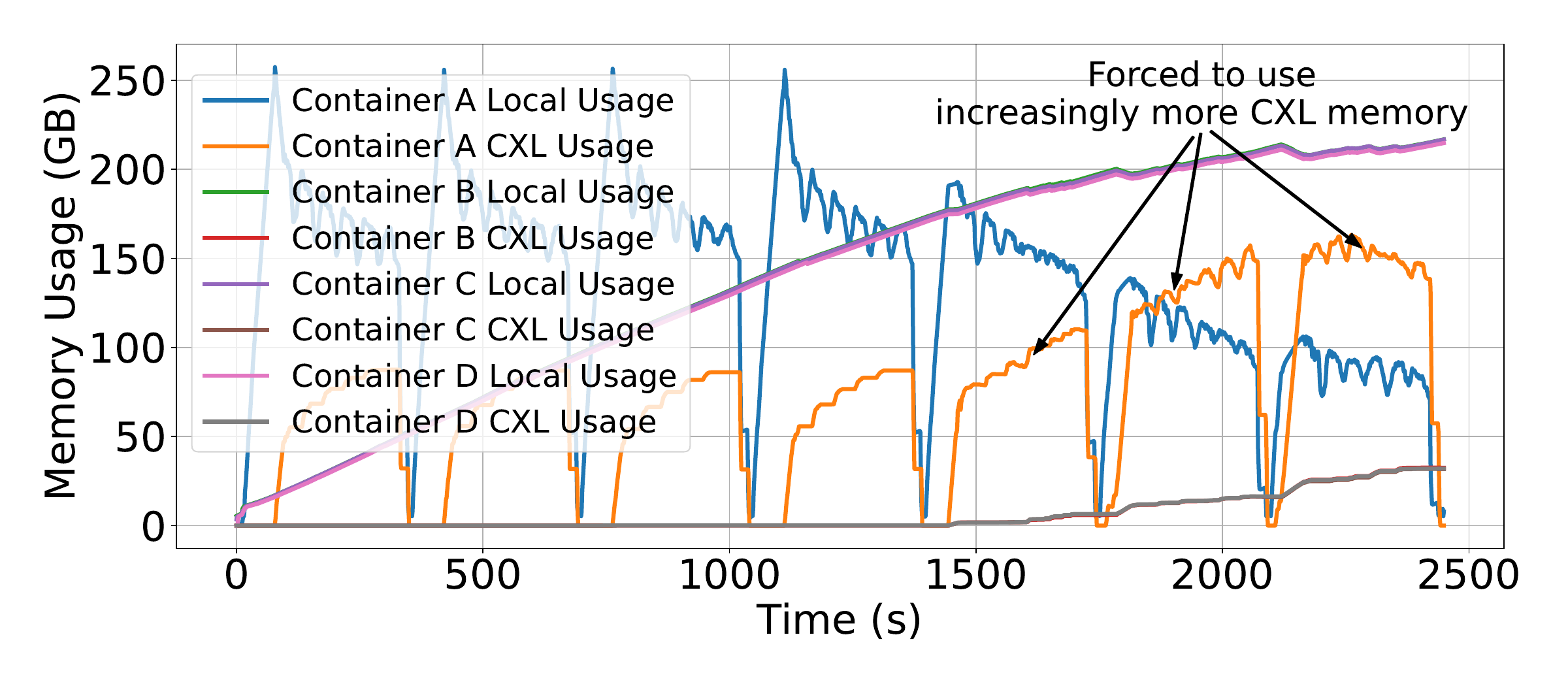}
\caption{Memory placement of TaoBench and SparkBench (Container A) on baseline Linux. Container A is forced to use increasingly more CXL memory and less local memory.}
\label{fig:dcperf-memory}
\vspace{-4mm}
\end{figure}

We select two workloads from DCPerf: SparkBench, which represents data analytics~\cite{spark}; and TaoBench, which represents a caching service~\cite{tao}.
In our experiment, we stack three instances of TaoBench (Containers B, C, and D) and one instance of SparkBench (Container A) per server.
TaoBench has a steady memory usage and access pattern, while SparkBench has bursty memory usage and varying access hotness over its data analytics phases.
Since SparkBench completes faster than TaoBench, we run SparkBench in a loop.
We set the local memory upper bound to be 192GB per container for \arch{}, which essentially divides the large server into four containers with equally sized shares of local memory.

We study the memory placement on baseline Linux as shown in Figure~\ref{fig:dcperf-memory}.
As TaoBench ramps up and consumes more memory, SparkBench with its less hot access patterns will be forced to use increasingly more CXL memory (orange line) and less local memory (blue line) in later iterations.
This leads to drastically slower performance for SparkBench.
We obtain on \arch{} 11.2 queries per hour (QPS) for SparkBench versus 4.2 QPS on Linux, which is a 1.7$\times$ throughput improvement.
The benefits come from the higher and more stable share of local memory that SparkBench is able to use when running under \arch{} as the maximum local memory per container is constrained.
Looking into memory placement with \arch{}, TaoBench is forced to spill all subsequent memory allocations past the 192GB upper bound into CXL memory, preserving free local memory for the short-lived and bursty SparkBench to use.
Notably, the local memory upper bound in \arch{} only marginally decreases SparkBench's performance by less than 0.3\%, demonstrating that applications have a wide range of sensitivity to memory placement and introducing fairness in \arch{} does not necessarily harm total system throughput.
This experiment demonstrates that \arch{} prevents the possible severe performance degradation when heterogeneous workloads are colocated.

\subsection{Deploying \arch{} in Production}
\label{sec:eval-case-studies}

We evaluate \arch{} with three case studies of deploying production workloads at \meta{}, demonstrating how \arch{} meets workload SLOs and improves workload performance with flexible fairness policies.
The three workloads are all memory-capacity-bound and have low CXL bandwidth utilization (\textless 10\% available bandwidth) in our experiments.

\subsubsection{Cache}

Cache exhibits random access patterns over its entire address space, with a significant portion of its memory footprint being hot (e.g., up to 60\% of the cache may be accessed concurrently). 
We stack two instances of Cache (Container A, B) on large servers, which fully saturates the application-usable memory capacity of the server. 
Because the two instances are homogeneous, the SLO calls for nearly identical cache look up latency. 

We first start both the containers at the same time on the baseline Linux without any fairness support. 
With both containers running homogeneous workloads, their memory usage across the tiers was supposed to be homogeneous.
However, with Linux, we see that for Container A, 90\% of its memory stays in local memory with the rest spilling to CXL memory, whereas Container B only keeps 70\% of its memory in local memory and uses significantly more CXL memory.
This unfair memory split across the memory tiers results in uneven application-level performance. 
More local memory helps Container A to serve more cache queries from the fastest memory tier:  
throughout the run, A has 1.15$\times$ better on average and as high as 3.3$\times$ better P99 latency over B. 

Although A is more advantageous in keeping an unfairly large amount of local memory when colocated with B under normal circumstances, without fairness support it can still fall victim to noisy neighbor effects.
To demonstrate this, we stop the loads to B, then artificially inject burst loads to B so that its memory utilization jumps from 0\% to 90\% within a minute.  
On Linux, rapid allocations by B forces A to demote 15\% of its anon memory to CXL tier and causes instability in A's performance. 
Cache being a latency sensitive workload, A experiences a huge spike in its tail latency and experiences a high failure rate -- request failures spike from 340 per minute to 83K per minute).
As a result, A's throughput drops by 65\%. 

We then run Cache on \arch{}, and we set the local memory lower protection to 70\% and upper bound to 75\% of the container's memory usage. 
We choose this configuration to resemble the server's local vs CXL memory ratio.
\arch{} is able to enforce the fair share of memory for the two containers.
In stable state, Container A and B maintain almost an identical memory usage across the tiers -- on average, 72\% and 73\% of their memory footprint is on local memory, respectively. 
With a more balanced local memory share, B's average P99 latency reduces by 47\% compared to Linux.
Meanwhile, A's latency also reduces by 52\% because of a 97\% reduction in NUMA hint faults -- setting the upper bound for both containers has limited the contention of local memory and every container spends less time on demotion and promotion that inflate application tail latency.
The result is that B has only 7\% higher average P99 latency over A, and for Cache, this uniform latency is key to meeting its SLO as an uneven latency in the object cache can propagate to other dependent services and cause cascading violations of performance guarantees in user-facing services.
This experiment shows that \arch{} helps colocated homogeneous workloads meet latency SLOs and avoid noisy neighbor effects.

\subsubsection{CI}

We evaluate collocated CI instances on large servers.
We stack four containers to execute four build jobs simultaneously, and compare runs on \arch{} and baseline Linux.
As the service is not customer-facing, it does not have a stringent SLO. 
Instead, the primary goal of \arch{} is to ensure that no containers are unfairly penalized due to local memory contention and to optimize for total job throughput.
We set on \arch{} the local memory lower protection to 192GB, which is derived by scaling down the expected memory usage of the container, 256GB, by the ratio of system local memory vs total memory.
This represents a simple fairness policy that does not need in-depth profiling of the workloads.

\begin{figure}[t]
\centering
\includegraphics[width=0.9\columnwidth, trim = 30mm 22mm 22mm 22mm]{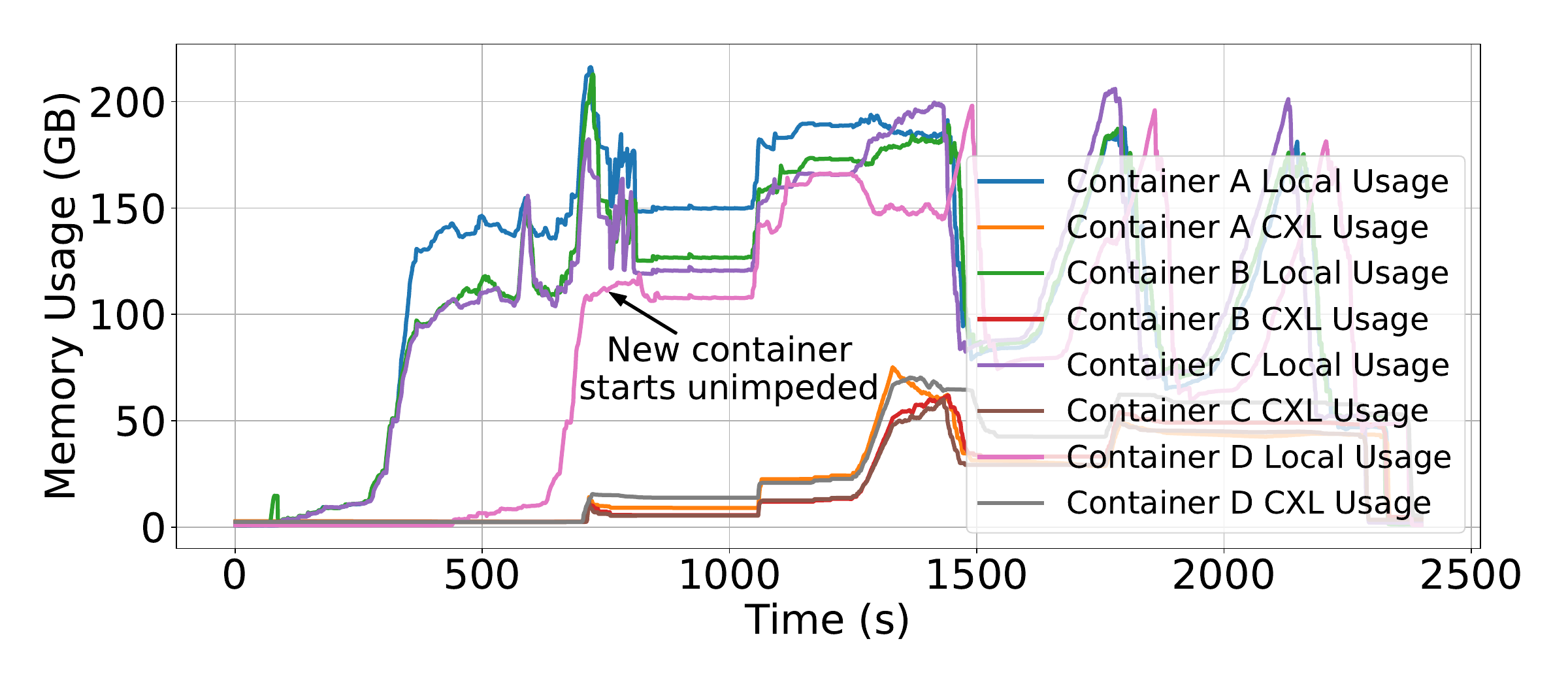}
\caption{Memory usage on \arch{} of four CI instances. }
\label{fig:sandcastle-memory}
\vspace{-4mm}
\end{figure}

CI has spiky memory capacity usage as depicted in Figure~\ref{fig:sandcastle-memory} showing its memory usage running on \arch{}, because linking phases of build jobs tend to be more memory-intensive than other phases.
In the absence of local memory pressure, Container A, B, C ramp up to freely utilize the available local memory capacity exceeding their fair share, leading to higher performance than stricter policies.
Then, shown in the figure from time 500s to 800s, Container D starts and rapidly allocates local memory, while other containers are forced to reduce their local memory footprint as they exceed their protected share.
We see that on \arch{}, Container D starts unimpeded and can put over 90\% of its memory footprint in local memory, whereas on Linux it is forced to put more than half of its footprint in CXL memory.
In terms of performance, \arch{} achieved a 3.3\% drop in average build time across the four build jobs relative to the Linux baseline. 
This experiment shows that with \arch{}, a simple policy derived from system memory capacity ratio is sufficient to improve total system throughput for some workloads in multi-tenant systems with CXL memory.

\subsubsection{Web}

We run colocated Web instances on small servers.
At runtime, an instances is specialized to serve a subset of traffic (partition) using just-in-time compilation. Therefore, different instances have different memory usage and access patterns despite sharing the same program binary.
Web is not memory capacity-bound on a per-instance level (i.e., increasing an instance's memory usage will not yield improved performance), and the gains of using CXL memory come from having more instances per server.
We consider three cases: stacked Web instances on \arch{}, on Linux baseline, and a single instance running alone using only local memory.
The performance of Web is measured as its average throughput per instance measured in requests per second (RPS) when internal latency limits are satisfied, and Web requires an SLO of less than 2\% slowdown compared to running alone using only local memory.
In the two stacked cases, we stack five instances of Web per server, with three instances belonging to partition A (Container B, C, E) and two instances belonging to partition B (Container A, D). 
We set on \arch{} the local memory lower protection to 28GB per container based on a profiling of the hot memory footprint of Web over a 10-minute period, ensuring there is sufficient local memory for an instance of Web to allocate its hot footprint without overcommitting the local memory.
The policy still allows some instances to utilize spare local memory capacity.

First, we compare end-to-end performance. 
On \arch{}, stacked Web instances do not lose performance compared to the single instance case, achieving 36 RPS for partition A and 80 RPS for partition B, meeting the service SLO.
On baseline Linux, although stacked instances serving partition A do not lose performance, instances serving partition B get only 74 RPS, missing the \textless 2\% performance loss SLO.
In other words, \arch{} allows 8\% more throughput for Web partition B than baseline running with multi-tenancy and CXL memory.

\begin{figure}[t]
\centering
\includegraphics[width=0.9\columnwidth, trim = 30mm 22mm 22mm 22mm]{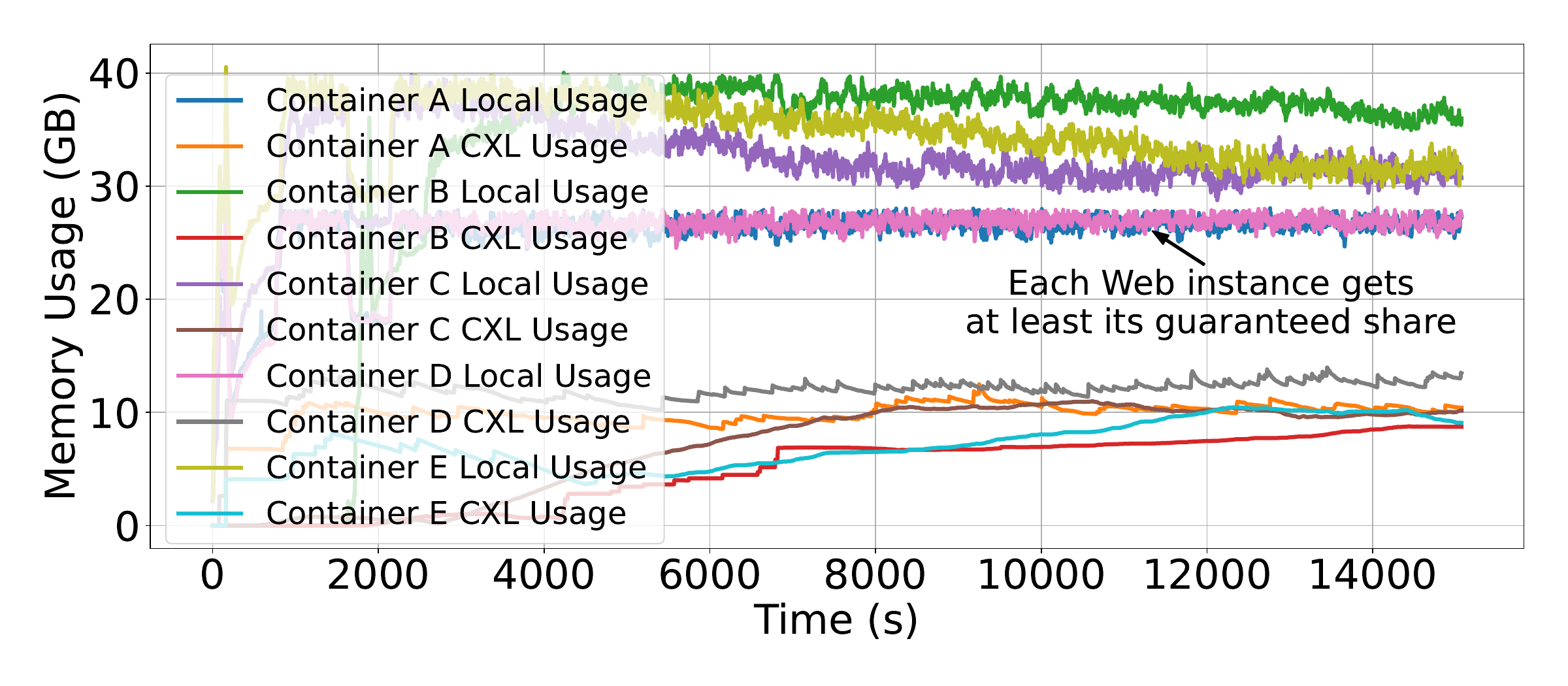}
\caption{Memory usage on \arch{} of five instances of Web.}
\label{fig:web-memory}
\vspace{-2mm}
\end{figure}

Next, we analyze the memory usage of colocated Web instances on baseline Linux. 
For Container A and D, we see their local memory usage trending downwards over time, decreasing by over 5GB over the course of the experiment.
This is because on Linux, their pages manifest as less hot than pages from other colocated Web instances and are gradually migrated to CXL memory.
The loss of local memory share for them leads to lower performance compared to \arch{}.

Contrast this with memory usage on \arch{} in Figure~\ref{fig:web-memory}.
Local memory lower protection is enforced, as every container is able to use at least 28GB of local memory. 
Note that Container A and D are using exactly 28GB of local memory at stable state, showing that \arch{} prevents instances on partition B from being forced to use an excessive amount of CXL memory as on Linux.
The experiments demonstrate that \arch{} can enforce the memory fair share policy for colocated containers and enable colocated performance that is close to running alone in systems with CXL memory.

\section{Related Work and Discussions}
\label{sec:related-work}

\begin{table}
    \centering
    \scriptsize
    \caption{Comparison of existing memory tiering designs.}
    \begin{tabular}{|l|c|c|c|c|}
        \hline
        Techniques & CXL & Upstream-First & Supports & Fairness \\
                    & Memory & Design & Multi-Tenancy & Policies \\
        \hline
        TPP~\cite{tpp} & $\checkmark$ & $\checkmark$ & X & X \\
        \hline
        Nimble~\cite{nimble} & X & X & X & X \\
        \hline
        Memstrata~\cite{flat2lm} & $\checkmark$ & X & $\checkmark$ & X \\
        \hline
        NOMAD~\cite{nomad} & $\checkmark$ & X & X & X \\
        \hline
        TMTS~\cite{tmts} & $\checkmark$ & X & $\checkmark$ & X \\
        \hline
        MEMTIS~\cite{memtis} & $\checkmark$ & X & X & X \\
        \hline
        Pond~\cite{pond} & $\checkmark$ & X & $\checkmark$ & X \\
        \hline
        Colloid~\cite{colloid} & $\checkmark$ & X & X & X \\
        \hline
        HeMem~\cite{hemem} & X & X & X & X \\
        \hline
        \textbf{\arch{}} & $\checkmark$ & $\checkmark$ & $\checkmark$ & $\checkmark$ \\
        \hline
    \end{tabular}
    \label{tab:prior-work}
    \vspace{-4mm}
\end{table}

Prior work~\cite{tpp,pond,memtis,tmts,alto,nimble,nomad,flat2lm,colloid,10032695,zhou2024neomem,sun2023demystifying,mtm,arif2024application,vtmm,thermostat,10.1145/3445814.3446713,10.1145/3695794.3695808,demeter,hemem,artmem} has built solutions that transparently managing memory placement of applications in the OS for systems with tiered memory.
We list a subset of prior work in Table~\ref{tab:prior-work} and categorize them based on (i) whether they are designed for CXL memory, which can differ significantly in performance characteristics from other tiered memory backends such as NVM, (ii) whether they are open-source and are designed with upstreaming to Linux kernel in mind, which is a prerequisite for maintainable deployment in datacenters in the long-term, (iii) whether they support multi-tenant setups, and (iv) whether they allow flexible fairness policies to support production workloads with diverse memory characteristics and performance requirements.

In terms of design objectives, prior work has primarily focused on optimizing page placement for the entire system.
However, fairness in multi-tenancy is important as colocated workloads require a certain level of performance predictability and guarantees regardless of their neighboring applications' behaviors. 
When prior work does address multi-tenancy, it tends to rely heavily on coarse-grained workload categories, such as \textit{critical/best-effort}~\cite{tmts,pond}.
This ignores the problem that colocated "critical" workloads need to meet their own performance SLOs but still suffer from noisy neighbor effects. In fact, at \meta{}, the vast majority of multi-tenancy setups are multiple instances of the same "critical" workload stacked together. 
In summary, flexible fairness policy support is critical for CXL memory to be viable in multi-tenant production environments.
Additionally, most prior work, except for TPP, is not upstreamed to Linux, making their production deployment difficult and long-term maintainability uncertain.

One prior work, TPP~\cite{tpp}, implements an OS-level memory tiering solution that builds on existing Linux memory management mechanisms. Specifically, TPP identifies cold pages to demote to CXL memory by leveraging Linux's page reclaim algorithm, which already tracks page hotness. For promotion, TPP utilizes Linux's NUMA balancing mechanism to identify hot pages in CXL memory.
TPP, however, lacks any notion of fairness and is not designed for multi-tenancy.

Another tiering work, MEMTIS~\cite{memtis} 
controls the ratio of the fast tier to capacity tier memory for its evaluation using the cgroup interface, supporting only an upper limit of the amount of fast tier memory that a cgroup can use, akin to allocating fast tier capacity to cgroups. 
\arch{}, in addition to supporting such upper limits, supports a much more powerful lower protection of the fast tier memory usage that is work-conserving and applied in relation to the local memory share of other colocated cgroups in the systems. Furthermore, demotions and promotions in \arch{} are regulated to conform to the lower protection, in contrast to the allocation-only enforcement in MEMTIS.
\arch{} is built and evaluated for multi-tenancy, while the cgroup limits in MEMTIS are only an aid to its evaluation.

\section{Conclusion}

\arch{} is a multi-tenant fairness solution for CXL memory tiering that provides per-container observability, simple yet powerful mechanisms for flexible fairness policies, and thrashing mitigation. \arch{} meets the performance SLOs of colocated workloads using CXL memory where TPP falls short, improves end-to-end performance of benchmarks and production workloads by up to 1.7$\times$ and 52\%, respectively, and is being deployed at \meta{} and upstreamed to Linux.

\bibliographystyle{IEEEtranS}
\bibliography{references}

\end{document}